\documentclass[prd,twocolumn,superscriptaddress]{revtex4}

\usepackage{graphicx}

\begin{document}

\title{The Role of Baryons in Unified Dark Matter Models}

\author{L.M.G. Be\c ca}
\email[Electronic address: ]{lmgb@fc.up.pt}
\affiliation{Centro de F\'{\i}sica do Porto e Departamento de F\'{\i}sica da Faculdade de
Ci\^encias da Universidade do Porto, Rua do Campo Alegre 687,
4169-007, Porto, Portugal}
\author{P.P. Avelino}
\email[Electronic address: ]{ppavelin@fc.up.pt}
\affiliation{Centro de F\'{\i}sica do Porto e Departamento de F\'{\i}sica da Faculdade de
Ci\^encias da Universidade do Porto, Rua do Campo Alegre 687,
4169-007, Porto, Portugal} \affiliation{Astronomy Centre,
University of Sussex, Brighton BN1 9QJ, United Kingdom}
\author{J.P.M. de Carvalho}
\email[Electronic address: ]{mauricio@astro.up.pt}
\affiliation{Centro de Astrof\'{\i}sica da Universidade do Porto,
R. das Estrelas s/n, 4150-762 Porto, Portugal}
\affiliation{Departamento de Matem\'atica Aplicada da Faculdade de
Ci\^encias da Universidade do Porto, Rua do Campo Alegre 687,
4169-007, Porto, Portugal}
\author{C.J.A.P. Martins}
\email[Electronic address: ]{C.J.A.P.Martins@damtp.cam.ac.uk}
\affiliation{Centro de Astrof\'{\i}sica da Universidade do Porto,
R. das Estrelas s/n, 4150-762 Porto, Portugal}
\affiliation{Department of Applied Mathematics and Theoretical
Physics, Centre for Mathematical Sciences,\\ University of
Cambridge, Wilberforce Road, Cambridge CB3 0WA, United Kingdom}
\affiliation{Institut d'Astrophysique de Paris, 98 bis Boulevard
Arago, 75014 Paris, France}

\begin{abstract}
We discuss the importance of including baryons in analyses of
unified dark matter scenarios, focusing on toy models involving a
generalized Chaplygin gas. We determine observational constraints
on this unified dark matter scenario coming from large scale
structure, type Ia Supernovae and CMB data showing how this
component can bring about a different behaviour from classical
$\Lambda$CDM and thus motivate further studies of this type of
models. We also speculate on interesting new features which are
likely to be important on non-linear scales in this context.
\end{abstract}
\date{24 March 2003}
\preprint{DAMTP-2003-26}
\maketitle

\section{Introduction}

Unified dark matter (UDM) models---also referred to as
`quartessence'---where dark matter and dark energy are seen as
different manifestations of a single substance have recently
enjoyed considerable attention, the main reason for it being, to
try and find viable alternatives to the highly fine-tuned potentials so
characteristic of quintessence models. The price one has to pay
seems to be the introduction of fluids obeying exotic equations of
state. One such example is the class of models based on the
Chaplygin gas, first suggested for this purpose in
\cite{Kamen,Bilic} and trivially generalized in \cite{Bento}.

There has been some controversy about the viability of these
scenarios. Following an earlier, simpler analysis \cite{Fabris},
we were able to obtain constraints on this type of model from a
combined dataset of 92 high-$z$ type Ia supernovae (including
forecasts for the proposed SNAP experiment) as well as the matter
power spectrum \cite{Avelino}. This has been subsequently repeated
by a number of authors \cite{Makler,Silva}, both of which confirm
our earlier results, while \cite{Colistete} have done a Bayesian
analysis of a sub-sample of the supernovae data set. The main
message of these studies is the fact that constraints critically
depend on whether one treats the Chaplygin gas as true
quartessence (replacing both dark matter and dark energy) or if
one allows it to coexist with a `normal' dark matter component
(which could be called the `Chaplygin quintessence' scenario). As
one might have expected \textit{ab initio}, the case for the
Chaplygin gas is stronger in the former scenario.

However, the above analyses essentially restrict themselves to the
background cosmology. Going beyond this and studying perturbation
theory in this context seemingly dealt the first blow to the UDM
possibility. The authors of \cite{Sandvik} have studied the
implications of a single generalized Chaplygin component to the
mass power spectrum and by comparing it to 2dF observations were
able to constrain parameters controlling it to such values as to
render it virtually indistinguishable from a usual $\Lambda$CDM
scenario. The production of violent oscillations causing the exponential
blowup of the power spectrum, totally inconsistent with
 observations was the reported fatal flaw. This was further
claimed to be of a general nature within UDM scenarios, which if
true could be construed as evidence for independent origins of
dark matter and dark energy.

The main goal of this Letter is to show that such flaw is in fact
due to \cite{Sandvik} overlooking a most important ingredient
describing the average Universe: \emph{baryons}. (This concern has
also been voiced, but not exploited, by \cite{Colistete}). Our
main objection can be brought down to the following point: The
main property (cosmologically speaking) of a Chaplygin gas is that
of mimicking cold dark matter (CDM) at early times as it
progressively evolves into a cosmological constant. This results
in perturbations in the Chaplygin gas component being heavily
damped at late times. However, if to it we add an independent
component with a low sound speed velocity (as baryons have), the
normal growth of inhomogeneities can still go on when the
Chaplygin gas starts behaving differently from CDM. By just
considering the Chaplygin fluid the authors of \cite{Sandvik} have
artificially constrained the possible power spectra that UDM
scenarios can cover. Although baryons are not that important for
other studies such as constraining these scenarios from supernova
luminosity distances, they are of critical importance in the context of large
scale structure studies.

We should also mention two recent papers which accurately
study perturbation growth in these models \cite{Finelli,Bean} (including
the cosmic microwave background). Although of a broader scope than
ours (they consider a \emph{baryon}$+$CDM$+$Chaplygin universe),
we do not quite agree on the latter's interpretation of UDM. While we
agree with their conclusion that this `Chaplygin quintessence'
scenario is all but ruled out (or at least disfavored by the
current data relative to $\Lambda$CDM) it seems to miss the point
that these scenarios came into existence as an attempt to unify
dark energy and dark matter, and that in this `quartessence'
context its behavior can be different from $\Lambda$CDM.
Yet this quartessence scenario is all but ignored in their discussion.
Therefore, we re-derived the analysis of \cite{Sandvik} (in order
to accommodate for baryons) and part of the analysis of
\cite{Bean} to explore more fully the viability of the
`quartessence' scenario.

\section{Growth of Perturbations}

To study the role of baryons in a UDM model based on the Chaplygin
gas we obviously need relativistic equations governing the
evolution of perturbations in a two fluid system. For a general
case of $n$ gravitationally interacting fluids, the linear
evolution of perturbations in the comoving synchronous gauge is
given by \cite{Shoba}:
\begin{eqnarray}
\label{eq1}
h'' + (2 + \xi )h' + 3\sum\nolimits_i {(1 + 3v_i^2 })\Omega_i\delta_i=0 \\
\label{eq2}
\delta '_i +(1+\omega_i)(\theta_i /a{\rm{H}} + h'/2)+ 3(v_i^2-\omega_i)\delta_i=0 \\
\label{eq3}
\theta '_i+(1-3v_i^2)\theta_i+ \frac{{v_i^2}}{{a{\rm{H}}(1+\omega_i)}}\nabla^2\delta_i=0
\end{eqnarray}
where $'\equiv d/dx$, $x=\ln a$ (note that \cite{Shoba} uses
conformal time instead), $h$ is the trace of the perturbation to
the Friedmann-Robertson-Walker (FRW) metric, H is the Hubble
parameter, $\xi=\rm{H}'/H$, $\delta_i$ is the density contrast of
the $i$th-fluid obeying $p_i=\omega_i\rho_i$ with an adiabatic
sound speed $v_i$ and a $\theta_i$ element velocity divergence.
Note that (\ref{eq2}) and (\ref{eq3}) apply for all
$i=1,\ldots,n$.

The first noticeable point is that for baryons (which we are
treating as an ordinary CDM fluid with $\omega_b=v_b^2=0$) Eq.
(\ref{eq3}) immediately renders $\theta_b=\theta_{b0}/a$.
Therefore, if  $\theta_{b0}=0$ then $\theta_b=0$ at all times.
Putting this into Eq. (\ref{eq2}) we find that $h' =-2\delta'_b$.
Taking this into consideration, the Fourier modes of a two fluid
model of baryons and a Chaplygin fluid with equation of state
$p=-C / \rho^\alpha$ (here $C$ is a positive constant and $\alpha
\ge 0$) turn out to be:
\begin{eqnarray}
\label{eq4}
\delta''_b+(2+\xi)\delta'_b-3/2\left[{\Omega_b\delta_b+(1-3\alpha\omega_{cg})\Omega_{cg}\delta_{cg}}\right]=0 \\
\label{eq5}
\delta'_{cg}+(1+\omega_{cg})\left[{\theta_{cg}/a{\rm{H}} -\delta'_b}\right]-3\omega_{cg}(1+\alpha)\delta_{cg}=0 \\
\label{eq6}
\theta'_{cg}+(1+3\alpha\omega_{cg})\theta_{cg}+ \frac{{\alpha\omega_{cg}k^2}} {{a{\rm{H}}(1+\omega_{cg})}}\delta_{cg}= 0
\end{eqnarray}
where we have used the fact that $v_{cg}^2=-\alpha\omega_{cg}$
\cite{Avelino} and $\nabla\equiv -k^2$. Thus we have three
equations for three unknowns: $\delta_b, \delta_{cg},
\theta_{cg}$. Given $\omega_{cg}$ and $\rm{H}$ (and $\xi$,
$\Omega_b$, $\Omega_{cg}$) as functions of $x$ we can easily
transform this set into four first order differential equations
and integrate it using a standard Runge-Kutta method. Since in the
linear regime and deep into the matter era $\delta_{b,cg} \propto
a$ implying $\delta_{b,cg}'\propto a$, normalized initial
conditions $[\delta_b, \delta_b', \delta_{cg},
\theta_{cg}]_0=[1,1,1,0]$ were used. As in \cite{Sandvik} we
evolve our system for a plane Universe from $z=100$ until today
and obtain corresponding transfer function $T_k$. But unlike in
the case of \cite{Sandvik}, our transfer function now comes from
the baryonic component and not from the Chaplygin component, thus
avoiding in the baryonic component the violent oscillations in the
Chaplygin gas that were so sensitive to $\alpha$.

Note that a (CDM processed) scale invariant Harrison-Zel'dovich
spectrum emerging from equality is accurately given by
\cite{Power}
\begin{equation}\label{ips}
\left|{\delta _k}\right|^2  =  A k
\left({\frac{\ln(1+\epsilon_0\zeta)}{\epsilon_0\zeta}}\right)^2 \left({
\sum_{i=0}^4(\epsilon_i \zeta)^i}\right)^{-1/2}
\end{equation}$\zeta=k/\Gamma h$, where $\Gamma=\Omega_{m0} h$
is the shape parameter ($\Omega_m^0$ being the CDM energy fraction
today), $A$, a normalization constant, $[k]={\rm{Mpc}}^{-1}$ and
$\epsilon=[2.34, 3.89, 16.1, 5.46, 6.71]$. As previously discussed
in \cite{Avelino} the presence of a Chaplygin fluid (still firmly
behaving as CDM at this epoch) only affects the shape parameter of
the power spectrum according to $\Gamma=\Omega^*_{m0}h$ where
\begin{equation}
\Omega^*_{m0} ={\Omega_{m0} + \Omega_{cg0}(1-\overline A)^{1/1+\alpha}},
\end{equation}
 $\alpha$ and $\overline A$ being the parameters controlling
the (generalized) Chaplygin gas. Unfortunately, this simple result
relies on the assumption of the absence of baryons. In order to
take them explicitly into account, Sugiyama's shape correction
\cite{Sugiyama} must be used instead
\begin{equation}\Gamma^*=\Omega^*_{m0} h \exp \left({-\Omega_{b0}(1+\sqrt{2h}/\Omega^*_{m0})} \right),\end{equation}
where $\Omega_{b0}$ stands for the baryon energy content today.
(Note that in the absence of baryons $\Gamma^*=\Gamma$ as it
should). Thus, we have taken (\ref{ips}) as our initial power
spectrum with a $\Gamma^*$ shape parameter, further processing it
through $T_k$ so as to compare it to 2dF results.

\section{The Analysis}

By performing a likelihood analysis of our model of baryons and a
Chaplygin fluid using the 2dF mass power spectrum we were able to
constrain the $(\alpha, \overline A)$ parameter space assuming
only reasonable priors coming from WMAP \cite{Spergel}, namely
$\Omega^0_b=0.044$ and $h=0.71$. Note that at the time of
recombination, the Chaplygin gas would still firmly behave as CDM.
Therefore, standard small scale CMB results are to be expected
when one identifies $\Omega_m$ with $\Omega^*_m$. The results
could conceivably differ on very large scales, though here they
would be competing against cosmic variance.

\begin{figure}[t]
\includegraphics[width=3.4in,keepaspectratio]{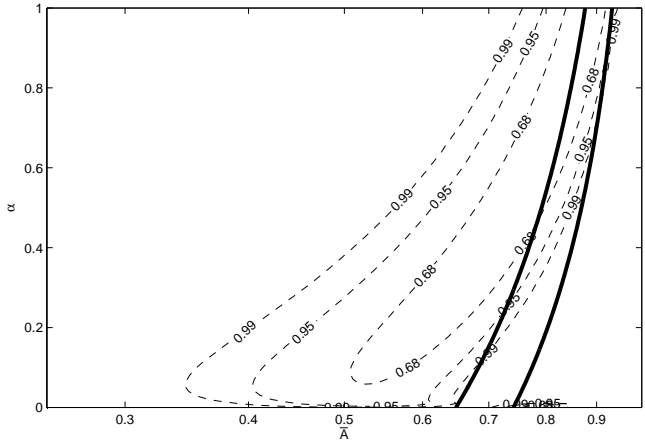}
\caption{$68\%$, $95\%$ and $99\%$ likelihood contours in the
$(\alpha,\overline A)$ parameter space for a model of baryons plus
a (generalized) Chaplygin gas, coming from the 2dF mass power
spectrum. Note the minute $\Lambda$CDM region near $\alpha \sim 0$
(see text). The zone inside the solid lines corresponds to
$\Gamma^*=0.2\pm0.03$ \cite{Dodelson}.} \label{contours}
\end{figure}

As in \cite{Sandvik} any 2dF data over $k>0.3~h{\rm{Mpc}}^{-1}$
was discarded so as to stay firmly grounded in the linear regime
where our analysis holds. Specifically, we have evaluated a
$500\times100\times100$ data grid for $A$, $\overline A$ and
$\alpha$ with $0<\alpha<1$ and $0<\overline A<1$, where
corresponding probabilities were found and posteriorly summed over
$A$. Resulting confidence levels are shown in Fig. \ref{contours},
where two disjoint regions can be found: one prominent, the other
small (around $\alpha \sim 0$ and $\overline A \sim 0.75$). We
have also displayed the region of parameter space corresponding to
a value of the shape parameter of $\Gamma^*=0.2\pm0.03$, as per
\cite{Dodelson}.

\begin{figure}[t]
\includegraphics[width=3.4in,keepaspectratio]{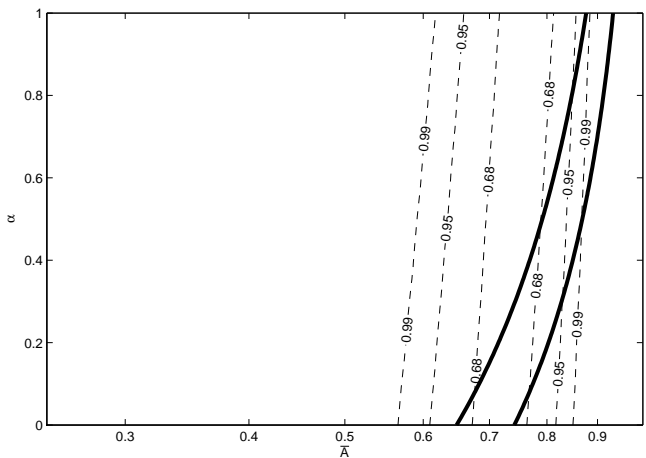}
\caption{$68\%$, $95\%$ and $99\%$ likelihood contours for a model
of baryons plus a (generalized) Chaplygin gas, coming from the 92
high-$z$ type Ia supernovae. The zone inside the solid lines
corresponds to $\Gamma^*=0.2\pm0.03$ \cite{Dodelson}.}
\label{supernovae}
\end{figure}

The significance of this smaller area is as follows. One of the
most noted proprieties of the Chaplygin gas class of models is
that of reproducing to all orders a classical $\Lambda$CDM
scenario when $\alpha=0$ (with $1-\overline A$ being the
equivalent CDM fraction $\Omega^*_{m0}$), which we know to be in
good agreement with observations, though still lacking a sound
fundamental physical motivation. In our case of baryons plus a
Chaplygin fluid, we are basically in the same $\Lambda$CDM
scenario, just with a slightly higher matter content because of
them. Therefore, this small area around $\alpha \sim 0$ is just
the $\Lambda$CDM component of the model manifesting itself. On the
other hand, we have an entirely disjoint region at very high
confidence level. So, in fact, we have provided explicit evidence
of a (generalized) Chaplygin gas not having to behave as
$\Lambda$CDM in order to reproduce 2dF large-scale structure data.

Now, let us re-visit high-$z$ type Ia supernovae. As previously
emphasized, these are only sensitive to the model background
evolution and not to the clustering details. So little information
is to be gained for $\alpha$ from a likelihood analysis of our
model of baryons plus a Chaplygin fluid for the supernova data on
their own. In fact, this analysis restricts mainly the value of
$\overline A$ since it more directly relates to the energy fueling
the background expansion---a point already made in \cite{Avelino}.
Fig. \ref{supernovae} shows the result of a likelihood analysis
using a combined dataset of 92 high-$z$ supernova observations
\cite{Perlmutter,Riess,Wang} (see \cite{Avelino,Pinto} for more
details).

\begin{figure}[td]
\includegraphics[width=3.4in,keepaspectratio]{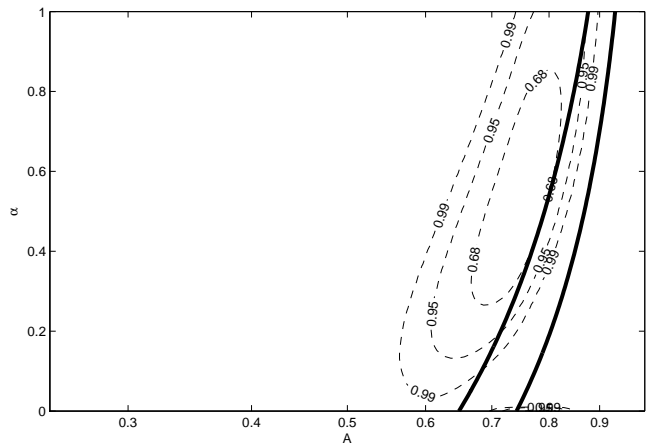}
\caption{$68\%$, $95\%$ and $99\%$ likelihood contours  resulting
from a joint analysis of large-scale structure (2dF) and type Ia
supernovae for a model of baryons plus a (generalized) Chaplygin
gas. The zone inside the solid lines corresponds to
$\Gamma^*=0.2\pm0.03$ \cite{Dodelson}.} \label{joint}
\end{figure}

However, notice that the degeneracy contours (large-scale
structure and supernovae) of the two observables, are not
parallel. This is of course to be expected since they are rather
different in nature and effectively arise from very different
redshifts. Therefore a joint analysis of Supernovae constraints
with those of large-scale structure can moderately improve the
constraints on $\alpha$, as shown in Fig. \ref{joint}.

\section{Conclusion}

We have shown that baryons play a crucial role in the context of
unified dark matter models involving a (generalized) Chaplygin
gas, their presence enabling them to reproduce the 2dF mass power
spectrum without having to behave as classical $\Lambda$CDM
scenarios The reason is that baryons can carry over gravitational
clustering when the Chaplygin fluid starts behaving differently
from CDM. This is of great importance since it allows the
(generalized) Chaplygin gas to be a conceivable quartessence
candidate.

On the other hand, it is easy to see why the Chaplygin
quintessence scenario will be observationally uninteresting, quite
apart from theoretical considerations: since in the standard
scenario there is currently no preference of the data for
quintessence over a plain cosmological constant, it is clear that
if we add a Chaplygin fluid to baryons and CDM then the available
data will prefer it to behave as a cosmological constant as
well---which is the result of \cite{Bean}.

Hence the present results, combined with those of \cite{Bean}
strengthen our previous conclusions \cite{Avelino}. If by an
independent method we determine the total matter density of the
universe to be $\Omega_m\sim0.3$ then in the context of this model
we would in fact \textit{require} a cosmological constant so as to
account for the current observational results. Conversely, the
quartessence case, where baryons are the only matter component
present, is such that its differences with respect to the standard
case are maximal, and indeed in this case the $\Lambda$-limit is
already strongly disfavored by observational data.

We finally wish to emphasize that this study, as well as the rest
of the published work so far, has been restricted to the linear
regime. Interesting new features of this scenario may also appear
on non-linear scales. An example would be the possibility of
avoiding the cuspy dark matter halo profiles expected in the
context of CDM models. We also anticipate that galaxy and cluster
evolution could be modified in the context of UDM models based on
the Chaplygin gas. This again results from the strong dependence
of the properties of the Chaplygin gas on the background density.

Moreover, in the non-linear regime the density is much higher than
the background density, and therefore one can not simply infer the
behavior of the Chaplygin fluid through the usual sound speed
(which uses the background density and pressure). So in a
collapsed region the Chaplygin gas could still behave as CDM even
though its background behavior may already be different. On the
other hand, since in this case the pertinent scales are very
small, even a tiny sound speed may be sufficient to halt collapse.
Clearly these issues must be addressed before one can reach a
final verdict on this scenario. We shall return to them in
subsequent work.

\begin{acknowledgments}
We thank Dominik Schwarz and the DAMTP Cosmology Journal Club for
useful comments and discussions. C.M. is funded by FCT (Portugal),
under grant no. FMRH/BPD/1600/2000. Additional support for this
project came from grant CERN/FIS/43737/2001.
\end{acknowledgments}

\bibliography{rescUDM}

\end{document}